# Electromagnetic PIC simulation with highly enhanced energy conservation

J. Yazdanpanah

*Department of Physics, Sharif University of Technology, P.O. Box 1155-4161, Tehran, Iran*

**Abstract**

We have obtained an electromagnetic PIC (EM-PIC) algorithm based on time-space-extended particle in cell model. In this model particles are shaped objects extended over time and space around Lagrangian markers. Sources carried by these particles are weighted completely into centers and faces of time-space cells of simulation-domain. Weighting method is resulted from implication of conservation of charge of shaped particles. By solving Maxwell's equations over source free zones of simulation grid we reduce solution of these equations to finding field values at nods of this grid. Major source of error in this model (and albeit other PIC models) is identified to be mismatching of particle marker location and location of its assigned sources in time and space. Relation of leapfrog scheme for integration of equations of motion with this discrepancy is investigated by evaluation of violation of energy conservation. We come in conclusion that instead of leapfrog we should integrate equations of motion simultaneously. Though equation of particle momentum becomes time implicit, we can solve it using a corrector-predictor method. In this way we obtain excellent improvement in energy conservation compared to existing leapfrog electromagnetic models. The developed theory is tested against results of our two dimensional EM-PIC code.

Electronic address: jamyazdanpanah@ mehr.sharif.ir , jamal.yazdan@ gmail.com

# I. Introduction

Investigation of highly nonlinear and unsteady phenomena in plasma physics demands for fully kinetic simulation methods. The Particle In Cell (PIC) method has proven itself to be a virtual kinetic lab for investigation of short space-time-scale phenomena such as interaction of high power short pulse lasers with plasma [1]. Possibility of parallel implementation of the method on currently available supercomputers and clusters has greatly facilitated study of sizable problems in this area of plasma physics.

General aspects of PIC algorithm are well described in refs [2], [3]. Variants of PIC schemes have been introduced [4-6]. Among them, well known Electromagnetic scheme assigns charge and currents into mesh nodes in a way that charge conservation is guaranteed and therefore eliminates need for solution of Gauss's law. This elimination results in highly local computational models that are easy to parallel implementation.

Accuracy of conservation of dynamical constants is measure of quality of PIC simulations. Each PIC scheme results in approximate forms for conservation laws. Multiplicity of effective factors in violation of exact conservations makes it hard to trace this violation back to these factors. Using objective models can greatly facilitate this difficulty because a connection is formed between these factors and they can bee treated with equal footing.

Here we proceed based on following picture of PIC; Particles are time-space-extended particles around Lagrangian markers as their centers. Both of time and space are included in simulation grid. Charge and currents of particles are completely weighted to centers and faces of cells (elements) of this grid. This weighting is generalization, to include explicit time dependence in particle shape function, of well known charge assignment of shaped particles in Hockney and Eastwood [3] p. 142. Conservation of total charge of a particle during its motion determines proper form of weighting of its sources. This conservation also results into an electromagnetic type algorithm. Inside source free zones of the grid the electromagnetic potentials are locally obtained in terms of their nodal values. This reduces problem of solution Maxwell's equations to finding nodal values of fields. Otherwise if we don't consider complete weighting, we can not evaluate fields self consistently with reasonable amount of computations.

Although after Lewis work [8], algorithms using method of variation of action integral have been named 'energy conserving', they don't conserve energy exactly. This is due to mismatch of energy exchange of particle and fields [7]. In our objective model this mismatch is easily identified to be a consequence of assignment of particle sources into time and space locations other than its marker location. After identification of relation of this discrepancy with time integration of equations of particle motion, we show that simultaneous integration of equations of motion instead of using common leapfrog method strongly suppresses consequences of this discrepancy and greatly enhances energy conservation.

Organization of paper is as follow. In the next section we obtain local profile of fields. In section (III) we discuss charge assignment and charge conservation of time dependently shaped particles. In section (IV) we discuss model's discrepancy. In section (V) Maxwell equations are integrated. In section (VI) we discuss old integration scheme of equations of motion and its consequences. In section (VII) a new scheme for integration of equations of motion is introduced. Finally we discuss results of PIC simulation based on this algorithm in sec (VIII).

In this paper we use following conventions and symbols; Nodal values are designated by labeled-capital-bold-alphabetic or Greek symbols, for example $\Phi_g^n$ and $\mathbf{A}_{i,g}^n$ are used respectively for scalar and vector potentials. Label of time level $n$ is as superscript and label of space $g$ is as subscript. If a vector quantity is considered, the first symbol in subscript, $i$, shows desired component. Sometimes we use direction name, for example $x$, to specify vector component. Mesh label $g$ may be multidimensional, $g = (l, m)$, or may (as well as time level) contain correctives such that $g = (l + 1/2, m)$. The same convention is used for assignment (interpolation) functions, in addition they contain continuous arguments for example as $w_{i,g}^n(t, x)$. Vectors (vector components and scalars) are designated by bold (simple) small alphabetic symbols. $\Delta_i$ is operator presentation of backward discrete differentiation such that $\Delta_1 V_g \equiv \Delta_x V_g = V_{l,m} - V_{l-1,m}$ for two dimensional mesh. In bounds of integrals we may use time (space) levels instead of time instant (position) itself, for example $\int_{n}^{n+1} .. \equiv \int_{t^n}^{t^{n+1}} ...$

## II. Local solution of fields

Consider particle markers to be located at $\{\mathbf{x}_p, \mathbf{v}_p\}$ in phase space. If we consider particles to be strongly time localized ($\delta$ shape in time) then charge and current densities $\rho$, $\mathbf{j}$ in time space are given by,

$$\rho(t, \mathbf{x}') = \sum_p Q_p S(\mathbf{x}' - \mathbf{x}_p(t)) \tag{2.1a}$$

$$\mathbf{j}(t, \mathbf{x}') = \sum_p Q_p S(\mathbf{x}' - \mathbf{x}_p(t)) \mathbf{v}_p(t) \tag{2.1b}$$

$S(\mathbf{x}' - \mathbf{x}_p(t))$ is shape function without explicit time dependence. Summation is taken over all particles and $Q_p$ is value of concentrated charge of particle $p$. In next section general explicitly time dependent shapes are considered, namely $S(t'-t, \mathbf{x}' - \mathbf{x}_p(t))$. We will show there that sources carried by particles are assigned to time-space grid nods $n$, $g$ according to time integrals over weight functions $u_g^n(t, \mathbf{x}_p)$ and $w_{i,g}^n(t, \mathbf{x}_p)$ for charge and $i$ th current respectively (see (3.15) and (3.16)). Demonstration is in sec (III) but expressions are given here,

$$\mathbf{Q}_g^n = \sum_p Q_p \int dt \, u_g^n(t, \mathbf{x}_p) \tag{2.2a}$$

$$\mathbf{J}_{i,g}^n = \sum_p Q_p \int dt [w_{i,g}^n(t, \mathbf{x}_p) v_i] \tag{2.2b}$$

$\mathbf{Q}_g^n$, $\mathbf{J}_{i,g}^n$ are nodal values of charge and current respectively. The complete weighting criteria implies following normalization condition for generalized shape function of particle,

$$\sum_{n,g} \int dt' \int d\tau' S(t'-t, \mathbf{x}' - \mathbf{x}_p(t)) \Big|_{\text{over time-space cell } (n,g)} = 1 \tag{2.3}$$

By complete weighting we can find local form of fields by solving the wave equation for potentials $\phi$ and $\mathbf{a}$ over source free time-space zones of computational grid. These potentials give electromagnetic fields $\mathbf{e}$, $\mathbf{b}$ via relations,

$$\mathbf{e} = -\nabla\phi - \frac{\partial \mathbf{a}}{\partial t} \tag{2.4a}$$

$$\mathbf{b} = \nabla \times \mathbf{a} \tag{2.4b}$$

For clarity assume a one dimensional wave equation governing scalar potential,

$$\frac{\partial^2 \phi}{\partial x^2} - \frac{1}{c^2}\frac{\partial^2 \phi}{\partial t^2} = 0 \tag{2.5}$$

Using method of separation of variables (see Jackson [9] p. 72) this equation can be solved over local zone with lower boundaries at $t = t^n$ and $x = x_l$ to give,

$$\phi(t,x) = (a_1 x + b_1)(a_2 t + b_2) + \sum_{n,l} c_l^n \sin(\frac{m(x-x_l)\pi}{Dx})\sin(\frac{k(t-t^n)\pi}{Dt}) + .. \tag{2.6}$$

Where $a_1, b_1, a_2, b_2$ and $c_l^n$ are constants that are determined by boundary conditions and $Dx$, $Dt$ are space and time differences. We discard all terms with wave numbers (frequencies) equal or larger than critical wave number $k_c$ (critical frequency $\omega_c$),

$$k_c = \frac{\pi}{Dx} \quad , \quad \omega_c = \frac{\pi}{Dt} \tag{2.7}$$

From equation (2.4a) for $\mathbf{e}$ field to be defined it is necessary that $\phi(t,x)$ be continuous in passage of space boundaries of grid elements,

$$\phi(t, x_l + \varepsilon) = \phi(t, x_l - \varepsilon) = \phi(t, x_l) \tag{2.8}$$

It is straightforward to show

$$\phi(t^n, x) = \Phi_l^n + \frac{\Phi_{l+1}^n - \Phi_l^n}{Dx}(x - x_l)$$

$$= (1 - \frac{|x - x_l|}{Dx})\Phi_l^n + (1 - \frac{|x - x_{l+1}|}{Dx})\Phi_{l+1}^n \tag{2.9}$$

Nodal values are boundary values $\Phi_l^n \equiv \phi(t^n, x_l)$. Continuity of $\phi(t,x)$ in passage of time boundaries is not expected generally. Using definition of triangle shape function $\Lambda_l(x)$ in (A2), more general expression for scalar potential in one dimension is written as,

$$\phi(t,x) = \sum_{n,l} f^n(t)\Lambda_l(x)\Phi_l^n \tag{2.10}$$

Where $f^n(t)$ is another local function that will be determined latter in this section. Now let consider a more general problem, two dimensional electromagnetic TM model; wave equations governing potential, in Lorenz gauge are written as,

$$\nabla^2\phi - \frac{1}{c^2}\frac{\partial^2\phi}{\partial t^2} = -\frac{\rho}{\varepsilon_0} \tag{2.11a}$$

$$\nabla^2\mathbf{a} - \frac{1}{c^2}\frac{\partial^2\mathbf{a}}{\partial t^2} = -\mu_0\mathbf{j} \tag{2.11b}$$

Conditions at boundaries of elements are,

$$\phi(t, x_l + \varepsilon, y) = \phi(t, x_l - \varepsilon, y) = \phi(t, x_l, y) \tag{2.12a}$$

$$\phi(t, x, y_m + \varepsilon) = \phi(t, x, y_m - \varepsilon) = \phi(t, x, y_m) \tag{2.12b}$$

$$a_x(t^n + \varepsilon, x, y) = a_x(t^n - \varepsilon, x, y) = a_x(t^n, x, y) \tag{2.12c}$$

$$a_y(t^n + \varepsilon, x, y) = a_y(t^n - \varepsilon, x, y) = a_y(t^n, x, y) \tag{2.12d}$$

$$a_x(t, x, y_m + \varepsilon) = a_x(t, x, y_m - \varepsilon) = a_x(t, x, y_m) \tag{2.12e}$$

$$a_y(t, x_l + \varepsilon, y) = a_y(t, x_l - \varepsilon, y) = a_y(t, x_l, y) \tag{2.12f}$$

The first two pairs come from (2.4a) and the third pair comes from (2.4b). Using these equations in the same way as what pursued to obtain (2.10), it is easy to show,

$$\phi(t, x, y) = \sum_{n,l,m} \Phi_{l,m}^n f^n(t)\Lambda_l(x)\Lambda_m(y) \tag{2.13a}$$

$$a_x(t, x, y) = \sum_{n,l,m} A_{x,l,m}^n \Lambda^n(t) g_l(x)\Lambda_m(y) \tag{2.13b}$$

$$a_y(t, x, y) = \sum_{n,l,m} A_{y,l,m}^n \Lambda^n(t)\Lambda_l(y) h_m(y) \tag{2.13c}$$

$g_l(x)$, $h_m(y)$ together with $f^n(t)$ are local functions that are not determined up to now. Note that undetermined parts of local functions correspond to directions that continuity of related potentials is not guaranteed. It is expected that extra restricting conditions limit undetermined parts of local functions. To see this we should use a result of section (IV) states that 'Lagrangian marker interpretation of particles (not energy conservation) results in equality of source weight functions and local profiles of corresponding potentials. In other words scalar potential $\phi$ and $i$ component of vector potential **a** are expanded in terms of their nodal values $\Phi_g^n$ and $\mathbf{A}_{i,g}^n$ according to,

$$\phi(t,\mathbf{x}) = \sum_{n,g} \Phi_g^n u_g^n(t,\mathbf{x}) \tag{2.14a}$$

$$a_i(t,\mathbf{x}) = \sum_{n,g} A_{i,g}^n w_{i,g}^n(t,\mathbf{x}) \tag{2.14b}$$

Therefore charge conservation affect form of local profile of potentials. In sec (III) we have shown that charge conservation implies that weight functions root in the same generating function $G_g^n$ (see relations (3.14)),

$$\Delta^t u_g^n(t,x) = -\frac{\partial G_g^n(t,x)}{\partial t} \tag{2.15a}$$

$$\Delta_i w_{i,g}^n(t,x) = -\frac{\partial G_g^n(t,x)}{\partial x_i} \tag{2.15b}$$

Here $G_g^n$ plays role of $S_g^n$ (here designated for shape function) local function in Eastwood [7]. Above relations can be written equivalently as,

$$\frac{\partial}{\partial t} \Delta_i w_{i,g}^n(t,x) = \frac{\partial}{\partial x_i} \Delta^t u_g^n(t,x) \tag{2.15c}$$

Substituting weight functions from (2.14) into this equation, it is easy to find undetermined functions $f^n(t)$, $g_l(x), h_m(y)$. After some straightforward manipulations, noting form of flattop function $\Pi$ in (A1), weight functions are completely determined as,

$$u_{l,m}^n(t,\mathbf{x}) = \Pi^n(t)\Lambda_l(x)\Lambda_m(y) \tag{2.16a}$$

$$w_{x,l,m}^n(t,\mathbf{x}) = \Lambda^{n+1/2}(t)\Lambda_{l+1/2}(x)\Lambda_m(y) \tag{2.16b}$$

$$w_{y,l,m}^n(t,\mathbf{x}) = \Lambda^{n+1/2}(t)\Lambda_l(x)\Pi_{m+1/2}(y) \tag{2.16c}$$

$$G_g^n(t,x) = \Lambda^{n+1/2}(t)\Lambda_l(x)\Lambda_m(y) \tag{2.16d}$$

We have used $1/2$ correctives into indices such that nodal values of scalar potential locate at integer levels. Above form for local functions satisfy the Lorenz gauge condition we have used to write wave equations in the form (2.11). It is easy to show within each local zone,

$$\nabla \cdot \mathbf{a} + \frac{1}{c^2}\frac{\partial \phi}{\partial t} = 0 \qquad (2.17)$$

Otherwise the above model wouldn't have such a solution.

Truncation of high frequency (wave number) part of expansion (2.6) is direct result of complete weighting. Error in local profile of potential, namely *local error*, due to this model is determined by comparing Taylor expansion of exact potentials with equations (2.13). For clarity let consider scalar potential in 1 dimension and eliminate time dependence. If $\psi$ be its exact form using $\psi(x_l) = \Phi_l$ and $\psi(x_{l+1}) = \Phi_{l+1}$ we will have in interval $[x_l, x_{l+1}]$,

$$\delta\phi_{err} = \psi(x) - \phi(x)$$

$$= \psi(x) - [\Phi_l + \frac{\Phi_{l+1} - \Phi_l}{Dx}(x - x_l)]$$

$$= -\frac{1}{2} Dx(x - x_l) \frac{\partial^2 \psi}{\partial x^2}\bigg|_{x=x_l} (1 - \frac{x - x_l}{Dx}) + h.o.t = O((\frac{Dx}{l_s})^2) \qquad (2.18).$$

$l_s$ is length scale of simulation, for example laser wavelength,…. The same statement holds for general situation.

### III. Source assignment and charge conservation

Here we discuss assignment of charge and currents of time dependently shaped particles. By linearity of Maxwell equations and continuity equation it is sufficient to consider effects of a solitary particle. Implication of conservation of charge of a shaped particle automatically gives convenient form of its assignment into the mesh.

Let begin with shape function without explicit time dependence (in form $S(\mathbf{x}' - \mathbf{x}_p(t))$) in one space dimension [3]. We can find exact values of electric field at boundaries of cells in context of our model. To do this we integrating Gauss's law at time $t'$ over cell $l$, we need to know value of total overlapped charge ($q_l(t')$) of particle $p$ with this cell ($\mathbf{e}(t', x_{l+1/2}) - \mathbf{e}(t', x_{l-1/2}) = q_l(t')$, see fig. (1a)),

$$q_l(t') = Q_p \int_{l-1/2}^{l+1/2} S(x' - x_p(t'))dx' = Q_p \int dt \int_{l-1/2}^{l+1/2} [\delta(t'-t)S(x' - x_p(t))]dx' \tag{3.1a}$$

For this case nodal fields are defined as $\mathbf{E}_{l+1/2}^n \equiv \mathbf{e}(t^n, x_{l+1/2})$, $\mathbf{E}_{l-1/2}^n \equiv \mathbf{e}(t^n, x_{l-1/2})$ nodal charge is defined respectively to be,

$$\mathbf{Q}_l^n \equiv q_l(t^n) \tag{3.1b}$$

We will have $\mathbf{E}_{l+1/2}^n - \mathbf{E}_{l-1/2}^n = \mathbf{Q}_l^n$. Such exact integration of Gauss's law confines local errors such as (2.19). Let consider variation of $q_l(t)$ due to infinite small displacement of particle from $\mathbf{x}_p$ to $\mathbf{x}_p + d\mathbf{x}_p$ is,

$$d(q_l(t)) = Q_p \int_{l-1/2}^{l+1/2} [S(x' - (x_p + dx_p)) - S(x' - x_p)]dx'$$

$$= -dx_p Q_p \int_{l-1/2}^{l+1/2} \frac{\partial S(x' - x_p)}{\partial x} dx' = -Q_p v_p dt[S(x_{l+1/2} - x_p) - S(x_{l-1/2} - x_p)] \tag{3.2}$$

Comparing right hand and left hand of the above equation, we come in conclusion that charge is accumulated at cell center by electric current $j$ with following form at surfaces of the cell,

$$j(t, x_{l+1/2}) = Q_p v_p S(x_{l+1/2} - x_p(t)) \tag{3.3}$$

Variation of charge during particle displacement in time interval $[t^n, t^{n+1}]$ is obtained by integration of (3.2). We integrate (3.2) to obtain,

$$\mathbf{Q}_l^{n+1} - \mathbf{Q}_l^n = -Q_p \int_n^{n+1} dt v_p [S(x_{l+1/2} - x_p) - S(x_{l-1/2} - x_p)] \tag{3.4}$$

Right hand side of (3.4) is net flux of particle charge trough space boundaries of the cell. Ampere's law encounters a time integral (to obtain electric field) in the same manner as space integration of Gauss's law. Therefore we define nodal value of current to be,

$$\mathbf{J}_{l+1/2}^{n+1/2} = Q_p \int_n^{n+1} dt v_p S(x_{l+1/2} - x_p(t)) = Q_p \int dt v_p(t) \int_n^{n+1} dt'[\delta(t'-t)S(x_{l+1/2} - x_p(t))] \tag{3.5}$$

Such assignment of charge and currents insures *conservation of charge* by (3.4). The rightmost expressions in (3.1a) and (3.5) give particle sources in terms of integrals over particle trajectory for strongly localized ($\delta$ shape) particle in time. Including time dependence in form factor decomposes time domain of particles and time domain of fields. To see this we can rewrite rightmost hands of equations (3.1a) and (3.5) respectively in form,

$$q_l(t') = Q_p \int_{l-1/2}^{l+1/2} dx' \int dt [\delta(t'-t) S(x'-x_p(t))] \qquad (3.6a)$$

$$\mathbf{J}_{l+1/2}^{n+1/2} = Q_p \int_{n}^{n+1} dt' \int dt [\delta(t'-t) v_p(t) S(x_{l+1/2}-x_p(t))] \qquad (3.6b)$$

Expressions inside $t$ integrals in the above equations are values of sources of a particle located at space-time point $(t, x_p(t))$ in space-time point $(t', x')$. In this way particle sources are gradually accumulated in point $(t', x')$ along with particle motion. Then sources in Maxwell's equations are integrated in the same way as charge in Gauss's law. To extend the above analysis we consider following generalization of shape function,

$$S(\mathbf{x}' - \mathbf{x}_p(t)) \to S(t'-t, \mathbf{x}' - \mathbf{x}_p(t)) \qquad (3.7)$$

Explicit time dependence states that a particle centered at point $(t, \mathbf{x}_p)$ along its world line is sensed at an arbitrary point of time and space $(t', \mathbf{x}')$. Consider amount of charge ($\Delta q_g^{n+1/2}(t)$) of a shaped particle included in a cell centered at time level $n+1/2$ and space level $g = (l, m)$ in analogy with (3.1),

$$\Delta q_g^{n+1/2}(t) = Q_p \int_{n}^{n+1} dt' \int_{g_-}^{g_+} d\tau' S(t'-t, \mathbf{x}' - \mathbf{x}_p(t)) \qquad (3.8)$$

With definition $g_- = (l-1/2, m-1/2)$ and $g_+ = (l+1/2, m+1/2)$. See fig.(1b) for a case in one space-dimension. When this particle moves along its word line, amount of its charge initially confined between time levels $t^n$ and $t^{n+1}$ gradually flows to its neighbor nods. Consider difference of $\Delta q_g^{n+1/2}(t)$ due to infinite small displacement of particle,

$$d(\Delta q_g^{n+1/2}(t)) = -dt Q_p \int_{n}^{n+1} dt' \int_{g_-}^{g_+} d\tau' [\frac{\partial S}{\partial t'} + \mathbf{v}_p \cdot \frac{\partial S}{\partial \mathbf{x}'}] \qquad (3.9)$$

It is useful to define *flux* in each direction to be integral of shape function over corresponding perpendicular hyper surface. Element of this hyper surface for time is $d\tau$ and for $x_i$ is multiplication of time difference $dt'$ and element of perpendicular usual surface $d\sigma_i'$, ($= dt' \cdot d\sigma_i'$). For two space dimensions ($x$ and $y$) we define,

$$F_t(t'-t, \mathbf{x}_g - \mathbf{x}_p(t)) \equiv \int_{g_-}^{g_+} d\tau' S(t'-t, \mathbf{x}' - \mathbf{x}_p(t)) \tag{3.10a}$$

$$F_x(t^{n+1/2} - t, x' - x_p, y_m - y_p) \equiv \int_n^{n+1} dt' \int_{m-1/2}^{m+1/2} dy' S(t'-t, \mathbf{x}' - \mathbf{x}_p(t)) \tag{3.10b}$$

$$F_y(t^{n+1/2} - t, x_l - x_p, y' - y_p) \equiv \int_n^{n+1} dt' \int_{l-1/2}^{l+1/2} dx' S(t'-t, \mathbf{x}' - \mathbf{x}_p(t)) \tag{3.10c}$$

We have used identity $S(t'-t, \mathbf{x}' - \mathbf{x}_p) \equiv S((t'-t^{n+1/2}) - (t - t^{n+1/2}), (\mathbf{x}' - \mathbf{x}_g) - (\mathbf{x}_p - \mathbf{x}_g))$ to determine arguments of densities. Manipulation of integral (3.9) in terms of these fluxes makes better insights in to their concept. For example the first term in integral of right hand side of (3.9) can be rewritten in the form,

$$\int_{t^-}^{t^+} dt' \int_n^{n+1} d\tau \frac{\partial S}{\partial t'} = \int_n^{n+1} dt' \frac{d}{dt'} \int_{g_-}^{g_+} d\tau S = F_t(t^+ - t, \mathbf{x}_g - \mathbf{x}_p(t)) - F_t(t^- - t, \mathbf{x}_g - \mathbf{x}_p(t))$$

Repeating the same straightforward operations for space differentiations, equation (3.9) becomes,

$$-d(\Delta q_g^{n+1/2}(t)) = Q_p \{dt[F_t(t^{n+1} - t, \mathbf{x}_g - \mathbf{x}_p(t)) - F_t(t^n - t, \mathbf{x}_g - \mathbf{x}_p(t))]$$

$$+ dx_p[F_x(t^{n+1/2} - t, \mathbf{x}_{l+1/2,m} - \mathbf{x}_p(t)) - F_x(t^{n+1/2} - t, \mathbf{x}_{l-1/2,m} - \mathbf{x}_p(t))]$$

$$+ dy_p[F_y(t^{n+1/2} - t, \mathbf{x}_{l,m+1/2} - \mathbf{x}_p(t)) - F_y(t^{n+1/2} - t, \mathbf{x}_{l,m-1/2} - \mathbf{x}_p(t))]\} \tag{3.11}$$

For a particle with finite extension and for arbitrary time level $n+1/2$ and space node $g$ we have,

$$\Delta q_g^{n+1/2}(-\infty) = \Delta q_g^{n+1/2}(\infty) = 0 \tag{3.12}$$

Note when we include time in particle coordinates it wouldn't be never at rest. Particle travels to neighborhood of time-space level from infinitely far distance and becomes far again as time proceeds. This implies charge conservation that is expressed in form,

$$\Delta q_g^{n+1/2}(\infty) - \Delta q_g^{n+1/2}(-\infty) = \int d(\Delta q_g^{n+1/2}(t)) = 0 \tag{3.13}$$

This is generalized form of charge conservation and may be interpreted as statement of 'no charge is left between time levels'. We can also rewrite other trivial relations between the fluxes and $\Delta q_g^{n+1/2}(t)$. Let define local generating function

$$G(t-t^{n+1/2}, \mathbf{x}-\mathbf{x}_g) \equiv G_g^{n+1/2}(t,\mathbf{x}) \equiv -\left.\frac{\Delta q_g^{n+1/2}(t)}{Q_p}\right|_{\mathbf{x}=\mathbf{x}_p(t)} \tag{3.14a}$$

Then from (3.11) it is clear that,

$$\Delta^t F_t = -\frac{\partial G}{\partial t} \tag{3.14b}$$

$$\Delta_i F_i = -\frac{\partial G}{\partial x_i} \tag{3.14c}$$

These relations are the same as what are used by Eastwood [7] to determine form of local functions ($G_g^{n+1/2}(t,\mathbf{x})$ here plays role of $S_g^{n+1/2}(t,\mathbf{x})$ there). But there reasoning was due to definiteness of electromagnetic potentials. Rewriting trajectory differences in terms of velocities ($dx_p = v_{p,x}(t)dt$, $dy_p = v_{p,y}(t)dt$) in (3.11), and integrating it over all times following forms for nodal values are neutrally obtained,

$$\mathbf{Q}_{l,m}^n = Q_p \int dt F_t(t^n - t, \mathbf{x}_{l,m} - \mathbf{x}_p(t)) \tag{3.15a}$$

$$\mathbf{J}_{x,l+1/2,m}^{n+1/2} = Q_p \int dt [v_{p,x}(t) F_x(t^{n+1/2} - t, \mathbf{x}_{l+1/2,m} - \mathbf{x}_p(t))] \tag{3.15b}$$

$$\mathbf{J}_{y,l,m+1/2}^{n+1/2} = Q_p \int dt [v_{p,y}(t) F_y(t^{n+1/2} - t, \mathbf{x}_{l,m+1/2} - \mathbf{x}_p(t))] \tag{3.15c}$$

Now it is easy to find connection between fluxes used here and weight functions of sec. II. That weight functions are exactly these fluxes,

$$u_g^n(t,\mathbf{x}) = F_t(t^n - t, \mathbf{x}_g - \mathbf{x}) \tag{3.16a}$$

$$w_{i,g}^{n+1/2}(t,\mathbf{x}) = F_i(t^{n+1/2} - t, \mathbf{x}_g - \mathbf{x}) \tag{3.16b}$$

As we have stated in descriptions of equations (3.6), eventually time domains of particles and fields are decomposed and we are left with cell integrated sources. To see this, let take a look at equations we have obtained so far. For example nodal charge as is given by (3.15a), is integration over cell space of following orbit integrated density,

$$\rho(t',\mathbf{x}') = Q_p \int dt S(t' - t, \mathbf{x}' - \mathbf{x}_p(t)) \tag{3.17a}$$

$$\mathbf{Q}_{l,m}^n = \int_{g-}^{g+} d\tau' \rho(t^n, \mathbf{x}') \tag{3.17b}$$

The same situation occurs about current densities and nodal currents,

$$\mathbf{j}(t',\mathbf{x}') = Q_p \int dt [\mathbf{v}_p(t) S(t' - t, \mathbf{x}' - \mathbf{x}_p(t))] \tag{3.18a}$$

$$\mathbf{J}_{x,l+1/2,m}^{n+1/2} = \int_{n}^{n+1} dt' \int_{m-1/2}^{m+1/2} d\sigma_x' \, j_x(t', x'_{l+1/2}, y') \tag{3.18b}$$

$$\mathbf{J}_{y,l,m+1/2}^{n+1/2} = \int_{n}^{n+1} dt' \int_{l-1/2}^{l+1/2} d\sigma_y' \, j_y(t', x', y'_{m+1/2}) \tag{3.18c}$$

Generalized form of charge conservation (3.11) that in terms of nodal values becomes,

$$\Delta^t \mathbf{Q}_g^{n+1} + \Delta_x \mathbf{J}_{x,l+1/2,m}^{n+1/2} + \Delta_y \mathbf{J}_{y,l,m+1/2}^{n+1/2} = 0 \tag{3.19}$$

## IV. Interaction of particles and fields and PIC discrepancy

It is well known that particle-field exchange term in electrodynamics is [9],

$$W_{ed} = \int dt \int d\tau [\rho\phi - \mathbf{a}.\mathbf{j}] \tag{4.1}$$

In analogy with this relation interaction of nodal sources and nodal fields is given by,

$$\mathbf{W} = \sum_{n,g} [\mathbf{Q}_g^n \Phi_g^n - \sum_i \mathbf{J}_{i,g}^n \mathbf{A}_{i,g}^n] \tag{4.2}$$

In the other hand we know particles as Lagrangian markers. This means that fields act on particles by interaction term,

$$\mathbf{W}' = \sum_p Q_p \int dt [\phi(t, \mathbf{x}_p) - \mathbf{A}(t, \mathbf{x}_p) \cdot \mathbf{v}_p] \tag{4.3}$$

For the whole system to be a Lagrangian system it is necessary that (4.2) and (4.3) be identically equal. Using relations (2.2) for nodal sources we end up with relations (2.4) for potentials. Basically we need to know form of variation of fields inside small local domains of mesh prior to computations. For this we have devised concept of complete assignment. In other hand particle is marked by a $2d$ dimensional vector in $d$ dimensional phase-space. Such particle-mesh model has an inevitable problem. To see this, let rewrite (4.3) in following useful form,

$$\mathbf{W}' = \sum_p Q_p \int d\tau \int dt [\delta(x - x_p(t))\phi(t, \mathbf{x}) - \delta(x - x_p(t))\mathbf{v}_p \cdot \mathbf{A}(t, \mathbf{x})]$$

It is clear that interaction term $\mathbf{W}'$ contains localized $\delta$ shape sources that possibly may be located inside infinite small local domains. While fields are calculated inside computational elements using edge and face centered sources, particles act upon the system by localized sources possibly inside elements. Note that this problem is inherent and we call it PIC discrepancy. A result of this discrepancy is violation of conservation laws. Because it is well known that conservation laws in fact result from symmetries in Lagrangian and any problem with a part of Lagrangian will affect these symmetries. It is neutral that decreasing size of the time space finite element reduces this discrepancy.

## V. Field equations for TM mode

As we have seen, in PIC source assignment is accompanied by a kind of accumulation in time-space (see equations (3.17) and (3.18)). Therefore to obtain variations of fields corresponding to these accumulated sources we should use integral form of Maxwell' equations. These equations are,

$$\oiint_S \mathbf{e}(t,\mathbf{x}') \cdot \mathbf{d\sigma} = \int_V \rho(t,\mathbf{x}') d\tau \qquad \text{(Gauss's law)} \qquad (5.1a)$$

$$\oint_L \mathbf{b}(t,\mathbf{x}') \cdot \mathbf{dl} = \frac{1}{c^2} \frac{d}{dt} \int_S \mathbf{e}(t,\mathbf{x}') \cdot \mathbf{d\sigma} + \frac{1}{c^2} \int_S \mathbf{j}(t,\mathbf{x}') \cdot \mathbf{d\sigma} \quad \text{(Ampere's law)} \qquad (5.1b)$$

$$\oint_L \mathbf{e}(t,\mathbf{x}') \cdot \mathbf{dl} = -\frac{d}{dt} \int_S \mathbf{b}(t,\mathbf{x}') \cdot \mathbf{d\sigma} \qquad \text{(Faraday's law)} \qquad (5.1c)$$

$$\oiint_S \mathbf{b}(t,\mathbf{x}') \cdot \mathbf{d\sigma} = 0 \qquad \text{(Absence of magnetic monopole)} \qquad (5.1d)$$

The Ampere's and Faraday's laws are integrated conveniently over time to give evolution of nodal fields. By expanding integrands in terms of nodal values of fields we obtain computational correspondences of these equations. To do this we should obtain local profiles of fields by substitution of expansions (2.14) into relations (2.4). For example using $e_x = -\partial \phi / \partial x - \partial a_x / \partial t$ we have,

$$e_x(t,x,y) = -\sum_{n,l,m} \frac{\Phi_{l,m}^n}{DX} \Pi^n(t) [-\Pi_{l+1/2}(x) + \Pi_{l-1/2}(x)] \Lambda_m(y) +$$

$$-\sum_{n,l,m} \frac{A_{x,l+1/2,m}^{n+1/2}}{DT} [-\Pi^{n+1}(t) + \Pi^n(t)] \Pi_{l+1/2}(x) \Lambda_m(y)$$

Changing summation index $l$ to $l+1$ in the second part of the first summation, and using local property of $\Pi_{l+1/2}(x)$ ( $\Pi_{l_0-1/2}(x) = \Pi_{l_1-1/2}(x) = 0$ where $l_0$ and $l_1$ are indices of points just before start and just after end of our computational domain ) it is easy to show,

$$\sum_{n,l,m} \Phi_{l,m}^n \Pi^n(t) \Pi_{l-1/2}(x) \Lambda_m(y) = \sum_{n,l,m} \Phi_{l+1,m}^n \Pi^n(t) \Pi_{l+1/2}(x) \Lambda_m(y)$$

Performing the same straightforward manipulations we have,

$$e_x(t,x,y) = \sum_{n,l,m} -(\frac{\Phi_{l+1,m}^n - \Phi_{l,m}^n}{DX} + \frac{A_{x,l+1/2,m}^{n+1/2} - A_{x,l+1/2,m}^{n-1/2}}{DT}) \Pi^n(t) \Pi_{l+1/2}(x) \Lambda_m(y)$$

$$\equiv \sum_{n,l,m} E_{x,l+1/2,m}^n \Pi^n(t) \Pi_{l+1/2}(x) \Lambda_m(y) \qquad (5.2a)$$

According to this relation, nodal values of $E_x$ locate at integer time, integer $y$ and half integer $x$ levels and relate to potential nodal values by,

$$E_{x,l+1/2,m}^n = -(\frac{\Phi_{l+1,m}^n - \Phi_{l,m}^n}{DX} + \frac{A_{x,l+1/2,m}^{n+1/2} - A_{x,l+1/2,m}^{n-1/2}}{DT}) \qquad (5.3a)$$

In the same way we obtain for other components and fields,

$$e_y(t,x,y) = \sum_{n,l,m} E_{y,l,m+1/2}^n \Pi^n(t) \Lambda_l(x) \Pi_{m+1/2}(y) \qquad (5.2b)$$

$$b_z(t,x,y) = \sum_{n,l,m} B_{z,l+1/2,m+1/2}^{n+1/2} \Lambda^{n+1/2}(t) \Pi_{l+1/2}(x) \Pi_{m+1/2}(y) \qquad (5.2c)$$

And

$$E_{y,l,m+1/2}^n = -(\frac{\Phi_{l,m+1}^n - \Phi_{l,m}^n}{DY} + \frac{A_{y,l,m+1/2}^{n+1/2} - A_{y,l,m+1/2}^{n-1/2}}{DT}) \qquad (5.3b)$$

$$B_{z,l+1/2,m+1/2}^{n+1/2} = -(\frac{A_{y,l+1,m+1/2}^{n+1/2} - A_{y,l,m+1/2}^{n+1/2}}{DX} - \frac{A_{x,l+1/2,m+1}^{n+1/2} - A_{x,l+1/2,m}^{n+1/2}}{DY}) \qquad (5.3c)$$

Mesh structure of fields and potentials is depicted in fig. 2.

Using nodal expansions obtained so far in equations (5.1), we obtain discreet Maxwell equations. Let examine Gauss's law first. Consider integral equation (5.1a) over a $2d$ volume determined in fig (3). We already have presented total charge inside the volume as nodal value of charge $\mathbf{Q}_{l,m}^n$. Therefore it is sufficient to calculate surface integral of the left hand side. Three nodal values of $\mathbf{e}$ contribute expansions (5.2a) and (5.2b) to determine values of this field over surface of each boundary. For example for the right boundary $e_x$ is determined in terms of $\mathbf{E}^n_{x,l+1/2,m-1}$ and $\mathbf{E}^n_{x,l+1/2,m}$ for points inside lower cell and $\mathbf{E}^n_{x,l+1/2,m}$ and $\mathbf{E}^n_{x,l+1/2,m+1}$ for points of upper cell. Integrals are calculated trivially to give

$$\Delta_x(\alpha_y \mathbf{E}^n_{x,l+1/2,m}) + \Delta_y(\alpha_x \mathbf{E}^n_{y,l,m+1/2}) = \mathbf{Q}^n_{l,m} \tag{5.4}$$

Where averaging operator $\alpha_l$ is defined such that $\alpha_l \mathbf{M}_l \equiv \frac{1}{8}\mathbf{M}_{l-1} + \frac{3}{4}\mathbf{M}_l + \frac{1}{8}\mathbf{M}_{l+1}$. As we have stated this equation is in fact an initial condition in satisfaction of charge conservation. Time evolution of electric field is governed by Ampere's law. Integrating this equation over time interval $[t^- = t^n, t^+ = t^{n+1}]$, we obtain,

$$\int_S \mathbf{e}(t^{n+1}, \mathbf{x}') \cdot \mathbf{d\sigma}' - \int_S \mathbf{e}(t^n, \mathbf{x}') \cdot \mathbf{d\sigma}' =$$

$$c^2 \int_n^{n+1} dt' \int_L \mathbf{b}(t, \mathbf{x}') \cdot \mathbf{dl}' - \int_n^{n+1} dt' \int_S \mathbf{j}(t, \mathbf{x}') \cdot \mathbf{d\sigma}' \tag{5.5}$$

Now consider the above integral over area enclosed by curve shown in fig (4b) that passes through points $\mathbf{x}_{l-1/2,m+1/2}$ and $\mathbf{x}_{l+1/2,m+1/2}$ tangential to $\mathbf{b}$ vector. Space bounds of this area together with time bounds are chosen such that current term in (5.5) coincides exactly with form of nodal current (3.18) obtained in sec (III). Again three neighbor nods contribute in electric flux through the above surface (see fig (4a)). The same situation occurs about time integration of $\mathbf{b}$ field (see fig (4c)). After fairly straightforward calculations we will have,

$$\Delta^t(\alpha_l \mathbf{E}^{n+1}_{y,l,m+1/2}) = c^2 \Delta_x(\alpha^t \mathbf{B}^{n+1/2}_{z,l+1/2,m+1/2}) - \mathbf{J}^{n+1/2}_{y,l,m+1/2} \tag{5.6a}$$

By choosing another convenient tangent curve passing through points $\mathbf{x}_{l+1/2,m-1/2}$ and $\mathbf{x}_{l+1/2,m+1/2}$, in the same manner we obtain Amperes law for $e_x$,

$$\Delta^t(\alpha_m \mathbf{E}^{n+1}_{x,l+1/2,m}) = -c^2 \Delta_y(\alpha^t \mathbf{B}^{n+1/2}_{z,l+1/2,m+1/2}) - \mathbf{J}^{n+1/2}_{x,l+1/2,m} \tag{5.6b}$$

To obtain computational form of Faraday's law we first time integrate this law equation over interval $[t^- = t^{n-1/2}, t^+ = t^{n+1/2}]$,

$$\int_S \mathbf{b}(t^{n+1/2}, \mathbf{x}') \cdot \mathbf{d\sigma}' - \int_S \mathbf{b}(t^{n-1/2}, \mathbf{x}') \cdot \mathbf{d\sigma}' = -\int_{n-1/2}^{n+1/2} dt' \int_L \mathbf{e}(t, \mathbf{x}') \cdot \mathbf{dl}' \tag{5.7}$$

To rewrite these integrals in terms of nodal values consider mesh presentation of fig (4d). We obtain straightforwardly,

$$\Delta^t \mathbf{B}_{z,l+1/2,m+1/2}^{n+1/2} = -(\Delta_x \mathbf{E}_{y,l,m+1/2}^n - \Delta_y \mathbf{E}_{x,l+1/2,m}^n) \tag{5.8}$$

When we work in regime of wave numbers (frequencies) that are very large comparing to length (time) differentials, as is the case in much of PIC simulations, a very cost effective work is to lump averaging operators in equations (5.4), (5.6a), (5.6b). Lumped versions of these equations are respectively,

$$\Delta_x \mathbf{E}_{x,l+1/2,m}^n + \Delta_y \mathbf{E}_{y,l,m+1/2}^n = \mathbf{Q}_{l,m}^n \tag{5.9a}$$

$$\Delta^t \mathbf{E}_{y,l,m+1/2}^{n+1} = c^2 \Delta_x \mathbf{B}_{z,l+1/2,m+1/2}^{n+1/2} - \mathbf{J}_{y,l,m+1/2}^{n+1/2} \tag{5.9b}$$

$$\Delta^t \mathbf{E}_{x,l+1/2,m}^{n+1} = -c^2 \Delta_y \mathbf{B}_{z,l+1/2,m+1/2}^{n+1/2} - \mathbf{J}_{x,l+1/2,m}^{n+1/2} \tag{5.9c}$$

## VI. Leapfrog scheme for explicit equations

Integral form of Newton-Lorenz equations of motion are,

$$\mathbf{p}(t^+) - \mathbf{p}(t^-) = \int_{t-}^{t+} dt [\mathbf{e}(t, \mathbf{x}_p(t)) + \mathbf{v}(t) \times \mathbf{b}(t, \mathbf{x}_p(t))] \tag{6.1a}$$

$$\mathbf{x}_p(t^+) - \mathbf{x}_p(t^-) = \int_{t-}^{t+} dt \mathbf{v}(t) \quad , \quad \mathbf{v} = \frac{\mathbf{p}}{\gamma} \quad , \quad \gamma = (1 - \mathbf{v} \cdot \mathbf{v})^{-1/2} \tag{6.1b}$$

Here after we eliminate particle index ($p$) of quantities except for position and use it only in summations over particles. Integrand of right hand of (6.1a) contains fields that are obtained from sources assigned from particles to mesh according to time integrals (3.18). A general computational strategy that previously was implicitly used is to approximate particle orbit piece by piece in intervals $[t^n, t^{n+1}]$s by linear interpolating functions. Within this approximation particle position is written in terms of its values at time instants $t^n$ s (see fig (5)),

$$x_{p,i}(t) = \sum_n \mathbf{X}_i^n \Lambda^n(t) \quad , \quad (\mathbf{X}_i^n \equiv x_{p,i}(t^n)) \tag{6.2}$$

Slop of interpolating function inside $[t^n, t^{n+1}]$ is defined to be velocity $\mathbf{V}^{n+1/2}$. With this definition, matching of interpolating line and exact orbit at bounds of this interval (see fig (5)) gives following exact relation for position update,

$$\mathbf{X}_i^{n+1} - \mathbf{X}_i^n = \mathbf{V}_i^{n+1/2} Dt \tag{6.3}$$

It is well known that with linear interpolation deviation of $\mathbf{V}^{n+1/2}$ from $\mathbf{v}(t^{n+1/2})$ is of order of $Dt^3$ (see ref [10] sec. 2) and maximum deviation of exact particle position from interpolating function at midpoint of $[t^n, t^{n+1}]$ is of order of $Dt^2$. If we use this approximate orbit inside integral (6.1a) its result will be accurate within (accuracy of $(\mathbf{x}_p) = Dt^2) \times Dt = Dt^3$. Therefore within accuracy of $Dt^3$ and using definition of $\mathbf{P}^{n+1/2} \equiv M\mathbf{V}^{n+1/2}(1-(\mathbf{V}^{n+1/2} \cdot \mathbf{V}^{n+1/2})/c^2)^{-1/2}$, Momentum equation is integrate over $[t^{n-1/2}, t^{n+1/2}]$ to give,

$$\mathbf{P}^{n+1/2} - \mathbf{P}^{n-1/2} = Q \int_{n-1/2}^{n+1/2} dt [\mathbf{e}(t, \mathbf{x}_p(t)) + \mathbf{v}(t) \times \mathbf{b}(t, \mathbf{x}_p(t))] \tag{6.4}$$

This equation should be simplified further to be implacable. Method of leapfrog replaces integrand of (6.4) with value of force at mid-time $t^n$. If electromagnetic fields were continuous and differentiable this replacement retains order of accuracy ($Dt^3$). This can be shown easily by Taylor expansion of integrand around $t^n$. Momentum update in Leapfrog scheme is,

$$\mathbf{P}^{n+1/2} - \mathbf{P}^{n-1/2} = QDt[\mathbf{e}(t^n, \mathbf{X}^n) + \mathbf{v}(t^n) \times \mathbf{b}(t^n, \mathbf{X}^n)] \tag{6.5}$$

According to (4.2c) $\mathbf{b}(t^n, \mathbf{X}^n) = \frac{1}{2}(\mathbf{b}(t^{n-1/2}, \mathbf{X}^n) + \mathbf{b}(t^{n+1/2}, \mathbf{X}^n))$ and we may write $\mathbf{v}_p(t^n) = \frac{1}{2}(\mathbf{V}^{n-1/2} + \mathbf{V}^{n+1/2})$ without loss of accuracy. But discontinuity of fields along some of time or space directions enters flaws into this approximation and reduces accuracy (see Birdsall and Langdon [2] section on effects of spatial grid).

Another inevitable approximation that we should make to obtain explicit algorithm is due to calculation of sources. If we try to obtain sources (3.17), (3.18) using constant speed approximation, time extension of particle relates these sources to values of dynamical variables at future times. To overcome this problem we should contract particle in time in source formulas. Form of generating function we have obtained in (2.16d) is due to a cubic shape particle in $2d$ space $+ 1d$ time. We calculate sources with strongly localized $\delta$ shape particle in time, with space smoothness is preserved. Such a change in shape function is equivalent to replacing of $\Lambda^{n+1/2}(t) \to \Pi^{n+1/2}(t)$ in generating function $G_{l,m}^{n+1/2}(t, \mathbf{x})$. Approximated weight functions that steel hold charge conservation are obtain from (2.15) to be,

$$u_{l,m}^n(t, x, y)(\text{inside charge integral}) \to \delta^n(t)\Lambda_l(x)\Lambda_m(y) \quad (6.6a)$$

$$w_{x,l+1/2,m}^{n+1/2}(t, x, y)(\text{inside current integral}) \to \Pi^{n+1/2}(t)\Pi_{l+1/2}(x)\Lambda_m(y) \quad (6.6b)$$

$$w_{y,l,m+1/2}^{n+1/2}(t, x, y)(\text{inside current integral}) \to \Pi^{n+1/2}(t)\Lambda_l(x)\Pi_{m+1/2}(y) \quad (6.6c)$$

Then source formulas (3.17), (3.18) reduce to,

$$\mathbf{Q}_{l,m}^n = \Lambda_l(\mathbf{X}_p^n)\Lambda_m(\mathbf{Y}_p^n) \quad (6.7a)$$

$$\mathbf{J}_{x,l+1/2,m}^{n+1/2} = \sum_p Q_p \mathbf{V}_{p,x}^{n+1/2} \int_n^{n+1} dt [\Pi_{l+1/2}(x_p(t))\Lambda_m(y_p(t))] \quad (6.7b)$$

$$\mathbf{J}_{y,l,m+1/2}^{n+1/2} = \sum_p Q_p \mathbf{V}_{p,y}^{n+1/2} \int_n^{n+1} dt [\Lambda_l(x_p(t))\Pi_{m+1/2}(y_p(t))] \quad (6.7c)$$

This approximation assigns sources exactly in the same way as refs [6], [7]. To realize its consequences, we should consider this approximation in association of using above constant speed approximation in time interval $[t^n, t^{n+1}]$. Because both of them contribute in PIC discrepancy discussed in sec (4). While fields are interacted by particles via source terms integrated over $[t^n, t^{n+1}]$, with constant speed approximation orbits of particles don't change in this interval. Generally as is seen from interaction formalism in sec (4), instantaneous interaction of particles and fields is violated during very small periods of time (time differences) or very small distances (space differences). Consequences of this problem can be better seen by examination of energy conservation. According to Poynting's theorem obtained for lumped equations (5.9) in APENDIX (B) we have,

$$\Delta^t U^{n+1} + \left( \mathcal{S}_{y,M}^{n+1/2} - \mathcal{S}_{y,0}^{n+1/2} + \mathcal{S}_{x,L}^{n+1/2} - \mathcal{S}_{x,L}^{n+1/2} \right) =$$

$$-\sum_{l,m} \left( \mathbf{J}_{x,l+1/2,m}^{n+1/2} \overline{\mathbf{E}}_{x,l+1/2,m}^{n+1/2} + \mathbf{J}_{y,l,m+1/2}^{n+1/2} \overline{\mathbf{E}}_{y,l,m+1/2}^{n+1/2} \right) \quad (6.7)$$

Symbols $U^{n+1}$, $S_{y,m}^{n+1/2}$ and $S_{x,l}^{n+1/2}$ are used for electromagnetic energy, electromagnetic flux along $y$ direction through boundary at $y = y_m$ and flux of energy along $x$ at $x = x_l$. And $\overline{\mathbf{E}}_{x,l+1/2,m}^{n+1/2} \equiv (\mathbf{E}_{x,l+1/2,m}^{n} + \mathbf{E}_{x,l+1/2,m}^{n+1})/2$. Right hand side gives rate of exchange of energy between fields and sources. We can obtain this exchange in other way by evaluation of rate of variation of kinetic energy of particles. Multiply equation of momentum update (6.5) by $\overline{\mathbf{P}}^n \equiv (\mathbf{P}^{n+1/2} + \mathbf{P}^{n-1/2})/2$, we obtain straightforwardly,

$$\gamma^{n+1/2} - \gamma^{n-1/2} = Q \frac{\mathbf{e}(\mathbf{X}^n) \cdot \overline{\mathbf{P}}^n}{\overline{\gamma}^n} \tag{6.8}$$

When right hand of above equation is compared with contribution of one particle in right of (6.7), they are unequal. Clearly form of currents also play role in this problem. Discontinuity of electric field in time (see equations (5.2a), (5.2b)) amplifies consequences of this mismatch. Smoothing shape of particle in time can be helpful to obtain better profiles for electric field but is hard to implement effectively.

We should solve these problems in self consistent manner. In next section we introduce an integration scheme that greatly suppresses these problems.

## VII. Simultaneous integration scheme

To suppress these problems, we should treat time integration more sophisticatedly. We perform integration of equations of motion (including momentum equation) and sources (3.19) simultaneously over time interval $[t^n, t^{n+1}]$ to incorporate simultaneous integration of sources and equations of motion. In the same way as previous section, orbit of particle is approximated piece by piece by interpolating lines but this time pieces are in intervals $[t^{n-1/2}, t^{n+1/2}]$ (see fig (5)). Slope of interpolating function inside interval $[t^{n-1/2}, t^{n+1/2}]$ is defined to be velocity $\mathbf{V}^n$ and its midpoint is defined to be $\mathbf{X}^n$. It is easy to show after little straightforward manipulations that,

$$\mathbf{X}^{n+1} - \mathbf{X}^n = \int_n^{n+1} \mathbf{v}_p(t)dt = \frac{Dt}{2}(\mathbf{V}^{n+1} + \mathbf{V}^n) \tag{7.1}$$

Also within linear interpolation accuracy $Dt^3$ equation of momentum is written as,

$$\mathbf{P}^{n+1} - \mathbf{P}^n = Q \int_n^{n+1} dt[\mathbf{e}(t,\mathbf{x}_p) + \mathbf{v}_p(t) \times \mathbf{b}(t,\mathbf{x}_p)] \tag{7.2}$$

We need further manipulations to recover form of force that causes exchange term of right hand of (6.7) reaches right hand of (6.8) as closely as possible. Let first consider translational part of force. For this, take a look at right hand side of Poynting's theorem (6.7). If we note carefully, we observe that with following changes we become very close to energy conserving force; first we should use inside integral (7.2) time average of interpolated orbit in time interval $[t^n, t^{n+1}]$ (see fig (5)),

$$\mathbf{x}_{av}(t) = \sum_n \Lambda^n(t)\mathbf{X}^n \tag{7.3}$$

Difference of above orbit with orbit introduced by (6.2) is that its nodal point doesn't match exactly with exact orbit at time instants $t^n$ s. As a result slop of this trajectory is given in terms of velocity to be $\frac{1}{2}(\mathbf{V}^{n+1} + \mathbf{V}^n)$ and we have,

$$\mathbf{x}_{av}(t)(\text{inside integral}) = \mathbf{X}^n + \frac{1}{2}(\mathbf{V}^{n+1} + \mathbf{V}^n)t \tag{7.4}$$

This replacement doesn't change level of accuracy of integral (7.4). Maximum deviation of averaged orbit and interpolating orbit is their difference at $t^{n+1/2}$ which is $|\mathbf{V}^{n+1} - \mathbf{V}^n|Dt/2 = O(Dt^2)$. Because interpolating orbit is match with exact orbit at $t^{n+1/2}$, deviation of zero orbit from exact trajectory also has the same order that result in $Dt^3$ accuracy of force integral.

The second change that we should do is to make $\mathbf{e}$ time continues and more concentrated inside interval $[t^n, t^{n+1}]$,

$$\mathbf{e}(t,\mathbf{x}_p)(\text{inside integral}) \equiv \left(\mathbf{e}(t^n,\mathbf{x}_p(t))\Pi^n(t) + \mathbf{e}(t^{n+1},\mathbf{x}_p(t))\Pi^{n+1}(t)\right)$$

$$\approx \frac{\left(\mathbf{e}(t^n,\mathbf{x}_p(t)) + \mathbf{e}(t^{n+1},\mathbf{x}_p(t))\right)}{2}\Pi^{n+1/2}(t) \tag{7.5}$$

Also with purpose of more easy application, we use following approximation for $\mathbf{b}$,

$$\mathbf{b}(t, \mathbf{x}_p)(\text{inside integral}) \equiv \mathbf{b}(t^{n+1/2}, \mathbf{x}_p)\Lambda^{n+1/2}(t) \approx \mathbf{b}(t^{n+1/2}, \mathbf{x}_p)\Pi^{n+1/2}(t) \quad (7.6)$$

Substituting these approximations in to (7.2) we obtain translational force $\mathbf{f}_e^{n+1/2}$,

$$f_{e,x}^{n+1/2} \equiv Q \int_n^{n+1} dt\, e_x(t, \mathbf{x}_p)$$

$$= Q \sum_{l,m} \overline{\mathbf{E}}_{x,l+1/2,m}^{n+1/2} \int_n^{n+1} dt [\Pi_{l+1/2}(x_p)\Lambda_m(y_p)] \quad (7.7a)$$

$$f_{e,y}^{n+1/2} \equiv Q \int_n^{n+1} dt\, e_y(t, \mathbf{x}_p)$$

$$= Q \sum_{l,m} \overline{\mathbf{E}}_{y,l,m+1/2}^{n+1/2} \int_n^{n+1} dt [\Lambda_l(x_p)\Pi_{m+1/2}(y_p)] \quad (7.7b)$$

Rotational force $\mathbf{f}_b^{n+1/2}$ is obtained to be,

$$\mathbf{f}_b^{n+1/2} \equiv Q_p \int_n^{n+1} dt\, [\mathbf{v}(t) \times \mathbf{b}(t, \mathbf{x}_p)]$$

$$= Q_p \frac{1}{2}(\mathbf{V}^{n+1} + \mathbf{V}^n) \times \sum_{l,m} \mathbf{B}_{l+1/2,m+1/2}^{n+1/2} \int_n^{n+1} dt [\Pi_{l+1/2}(x_p)\Pi_{m+1/2}(y_p)] \quad (7.7c)$$

And equation of motion (7.2) reads as,

$$\mathbf{P}^{n+1} - \mathbf{P}^n = \mathbf{f}_e + \mathbf{f}_b \quad (7.8)$$

Using approximation of average orbit (7.4) into current integrals (2.2) and using relations (6.6) gives form of currents,

$$\mathbf{J}_{x,l+1/2,m}^{n+1/2} = \frac{1}{2} \sum_p Q_p (\mathbf{V}_{p,x}^{n+1} + \mathbf{V}_{p,x}^n) \int_n^{n+1} dt [\Pi_{l+1/2}(x_p(t))\Lambda_m(y_p(t))] \quad (7.9a)$$

$$\mathbf{J}_{y,l,m+1/2}^{n+1/2} = \frac{1}{2} \sum_p Q_p (\mathbf{V}_{p,y}^{n+1} + \mathbf{V}_{p,y}^n) \int_n^{n+1} dt [\Lambda_l(x_p(t))\Pi_{m+1/2}(y_p(t))] \quad (7.9b)$$

Then analogous expression to (6.8) is obtained by inner production of equations of motion (7.10) by $\overline{\mathbf{P}}^{n+1/2} \equiv (\mathbf{P}^n + \mathbf{P}^{n+1})/2$,

$$\gamma^{n+1} - \gamma^n = Q \frac{\mathbf{f}_e^{n+1/2} \cdot \overline{\mathbf{P}}^{n+1/2}}{\overline{\gamma}^{n+1/2}} \quad (7.10)$$

Comparing right hands of (7.10) and (6.7) with new currents (7.9) we conclude that two expressions match for one particle contribution within accuracy of,

$$\frac{\mathbf{P}^{n+1}}{\gamma^{n+1}}+\frac{\mathbf{P}^{n}}{\gamma^{n}}-\frac{\overline{\mathbf{P}}^{n+1/2}}{\overline{\gamma}^{n+1/2}} \approx -\frac{(\gamma^{n+1}-\overline{\gamma}^{n+1/2})}{(\overline{\gamma}^{n+1/2})^{2}}(\mathbf{P}^{n+1}-\overline{\mathbf{P}}^{n+1/2}) \qquad (7.11)$$

That is second order accurate in terms of time-continuous variables. The only remained problem is implicit form of forces (7.7). At time of evaluation of forces we don't have trajectory of particle. At this stage we can use a corrector-predictor method. Value of displacement of particle is taken to be equal to its value at previous time step. Because particle displacement is a continuous function of time the resulted error ($Dt^2$ in force integral) is much less than error due to ill defined forces in right hand of (6.5). We will discuss results of this algorithm in the next section.

## VIII. Results and discussions

We have written a 2D EM-PIC code including both of our model and leapfrog model as its options. For our model, it uses equations (5.9) for field update and relations (7.9) for source assignment and equations (7.8) for particle update. For the other model, it uses equations (6.3), (6.5) instead of (7.8) and assigns sources by (6.7). It is a C implementation of discussed EM algorithm that uses Message Passing Interface (MPI) for parallelization. Boundary conditions for fields are periodic for upper and lower boundaries and free space for side boundaries. It uses periodic boundary conditions for up-crossing and down-crossing particles and completely reflecting boundary conditions for side crossing particles. For demonstration of correctness of procedure of implementation of algorithm, we have visualized in fig (6), results of interaction of a $\tau = 60\,fs$ laser having normalized vector potential $a_0 = 16$ and spot diameter $r_\perp = 3\mu m$ at wavelength $\lambda_0 = 1\mu m$, with near critical plasma having initial density of $n_0/n_{cr} = 0.9$. Dimensions of plasma and simulation box are shown in figures.

In order to demonstrate our analysis and our model advantages we have compare energy-conservation of our 'simultaneous time integration' scheme with commonly used leapfrog scheme. Percentage of absorbed energy respect to portion of laser energy inside simulation box is measured in two ways, first by evaluation of variations of field energy and second by evaluation of net variation of energy of particles. History of this absorption is given in fig (7), for $N_{pc} = 0.5$ and $N_{pc} = 5$ particles per grid cell.

Parameters of laser and plasma are the same as previous parameters except for laser amplitude $a_0 = 2$ that chosen such that thermal energy of particles is noticeable relative to electromagnetic energy. Target thickness is chosen to be $L = 2\mu m$, making it possible to observe both heating and cooling of plasma. Temperature of electrons (ions) is $T_e = 1 keV$ ($T_i = 0$). Computational parameters are $Dx = \lambda_0/119$, $Dy = \lambda_0/24$ and $Dt = 0.5 \times \left(Dt_{max} = (1/c)(1/Dx^2 + 1/Dy^2)^{-1/2}\right)$. At very low number of particles that average distance of them becomes comparable to neutral scales of plasma, interaction of errors with these modes become more effective and results in strong violation of energy conservation as can be seen in results of leapfrog model (see fig (7c)). Note excellent improvement of energy conservation in results of our model especially at lower particle number $N_{pc} = 0.5$. Increasing number of particles per cell controls grow up of numerical errors as pointed out previously [2], [3]. A noticeable point is that generally value of absorption increased with increment of particle number irrespective of energy conservation (compare value of absorption for different particle number) and ultimately will be saturated. We aim to study this effect carefully in our future investigations.

In conclusion we have presented a detailed study on time-space-extended particle in cell model. Sophisticated analysis of this model has resulted in a new electromagnetic PIC algorithm with corrected equations of motion and corrected source assignment for enhancement of energy conservation. Results of implementation of this algorithm on computer, excellently confirms our analysis.

## APENDIX A: Definitions of local functions

Definitions of local functions are,

$$\Pi_l(x) \equiv \Pi(x-x_l) \equiv \begin{cases} 1 & |x-x_l| \leq Dx/2 \\ 0 & elsewhere \end{cases} \qquad (1)$$

$$\Lambda_l(x) \equiv \Lambda(x-x_l) \equiv \begin{cases} 1 - \dfrac{|x-x_l|}{Dx} & |x-x_l| \leq Dx \\ 0 & elsewhere \end{cases} \qquad (2)$$

## APENDIX B: Poynting's theorem

To obtain Poynting's theorem let first obtain computational correspondence of part by part integration. Consider two indexed quantities $Z_1^n$ and $Z_2^n$, with index $n$ ranged from $0....N$. Then neutrally backward time difference of these quantities is defined for $n \geq 1$. Discrete form of part by part integration is,

$$\sum_n Z_2^n \Delta^n Z_1^n = -\sum_n Z_1^{n-1} \Delta^n Z_2^n + \left(Z_1^N Z_2^N - Z_1^0 Z_2^0\right) \tag{1}$$

This is only a rearrangement of summands inside summations. Now multiply equation (5.9c) by $\overline{\mathbf{E}}_{x,l+1/2,m}^{n+1/2} \equiv \frac{1}{2}(\mathbf{E}_{x,l+1/2,m}^n + \mathbf{E}_{x,l+1/2,m}^{n+1})$ and equation (5.9b) by $\overline{\mathbf{E}}_{y,l,m+1/2}^{n+1/2} \equiv \frac{1}{2}(\mathbf{E}_{y,l,m+1/2}^n + \mathbf{E}_{y,l,m+1/2}^{n+1})$, and sum over $0 \leq l \leq L$ and $0 \leq m \leq M$ then sum up the results, after tedious but fairly straightforward calculations we obtain Poynting's theorem. For example from the first multiplication we obtain expressions like $\sum_{l,m} \mathbf{E}_{x,l+1/2,m}^n \Delta_y \mathbf{B}_{z,l+1/2,m}^{n+1/2}$ that using (1) transforms into,

$$\sum_{l,m} \mathbf{E}_{x,l+1/2,m}^n \Delta_y \mathbf{B}_{z,l+1/2,m}^{n+1/2} =$$
$$-\sum_{l,m} \Delta_y \mathbf{E}_{x,l+1/2,m}^n \mathbf{B}_{z,l+1/2,m-1/2}^{n+1/2} + \sum_l \left(\mathbf{E}_{x,l+1/2,M}^n \mathbf{B}_{z,l+1/2,M+1/2}^{n+1/2} - \mathbf{E}_{x,l+1/2,0}^n \mathbf{B}_{z,l+1/2,1/2}^{n+1/2}\right) \tag{2}$$

The first term in the right hand side is a part of right hand of Faraday's law multiplied by $\mathbf{B}_{z,l+1/2,m-1/2}^{n+1/2}$ that in addition with the similar terms from the other component of electric field give $\mathbf{B}_{z,l+1/2,m+1/2}^{n+1/2} \Delta^t \mathbf{B}_{z,l+1/2,m+1/2}^{n+1/2}$, and the other terms in right hand are parts of surface summation over Poynting's vector. We will end up with,

$$\Delta^t U^{n+1} + \left(\mathcal{S}_{y,M}^{n+1/2} - \mathcal{S}_{y,0}^{n+1/2} + \mathcal{S}_{x,L}^{n+1/2} - \mathcal{S}_{x,L}^{n+1/2}\right) =$$
$$-\sum_{l,m} \left(\mathbf{J}_{x,l+1/2,m}^{n+1/2} \overline{\mathbf{E}}_{x,l+1/2,m}^{n+1/2} + \mathbf{J}_{y,l,m+1/2}^{n+1/2} \overline{\mathbf{E}}_{y,l,m+1/2}^{n+1/2}\right) \tag{3}$$

With following definitions of electromagnet energy ($U^n$) and Poynting's flux components ($\mathcal{S}_{x,l}^{n+1/2}$, $\mathcal{S}_{y,m}^{n+1/2}$),

$$U^n = \frac{1}{2} \sum_{l,m} \left((\mathbf{E}_{x,l+1/2,m}^n)^2 + (\mathbf{E}_{y,l,m+1/2}^n)^2 + \mathbf{B}_{z,l+1/2,m+1/2}^{n+1/2} \mathbf{B}_{z,l+1/2,m+1/2}^{n-1/2}\right)$$

$$\mathcal{S}_{x,l}^{n+1/2} = \sum_m \overline{\mathbf{E}}_{y,l,m+1/2}^{n+1/2} \mathbf{B}_{z,l+1/2,m+1/2}^{n+1/2}$$

$$\mathcal{S}_{y,m}^{n+1/2} = -\sum_l \overline{\mathbf{E}}_{x,l+1/2,m}^{n+1/2} \mathbf{B}_{z,l+1/2,m+1/2}^{n+1/2}$$

Don't conflict symbol of Poynting's vector with shape function and symbol of energy with charge weighting function.

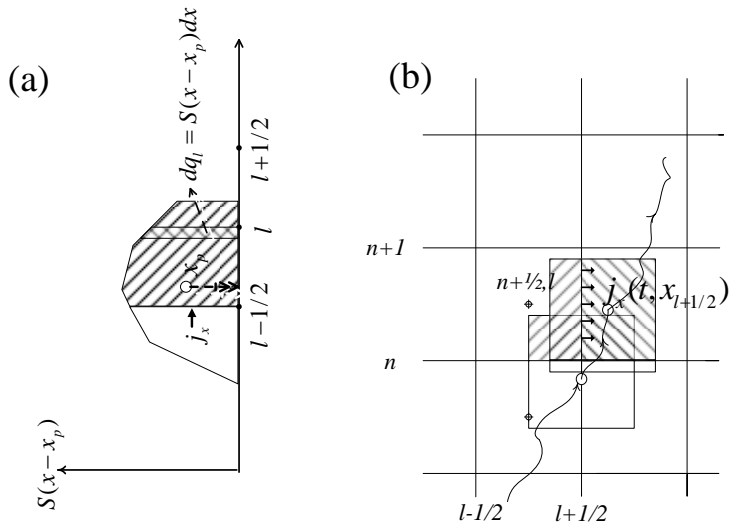

**Figure 1: (a) Assignment of charge of strongly time-localized shaped particle. Portion of charge located inside cell boundaries, has been assigned to cell center. This charge has been accumulated gradually by currents through cell boundaries. (b) When a time extended particle moves along its word line in the same way its charge gradually is accumulated into a cell.**

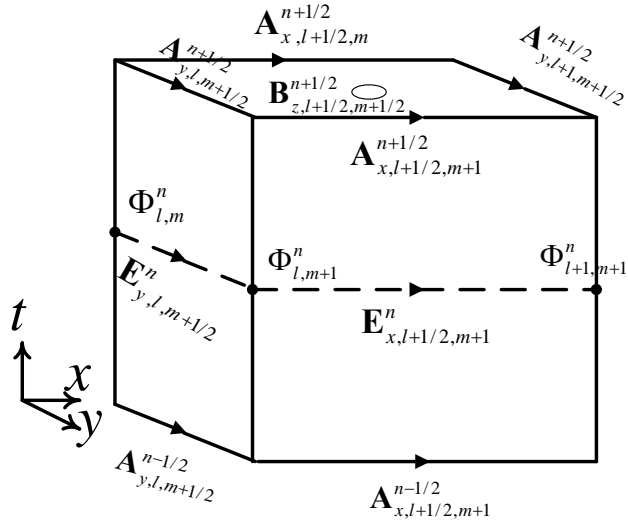

**Figure 2: Structure of time-space element.**

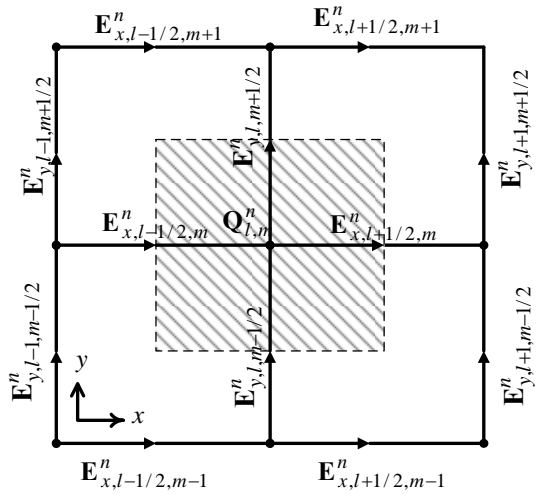

**Figure 3: x-y layout of four adjacent domain elements at $t = t^n$. Nodal values of $e$ are presented. Gauss law is integrated over shadowed volume. The total charge inside this volume $Q_{l,m}^n$ is assigned to central point.**

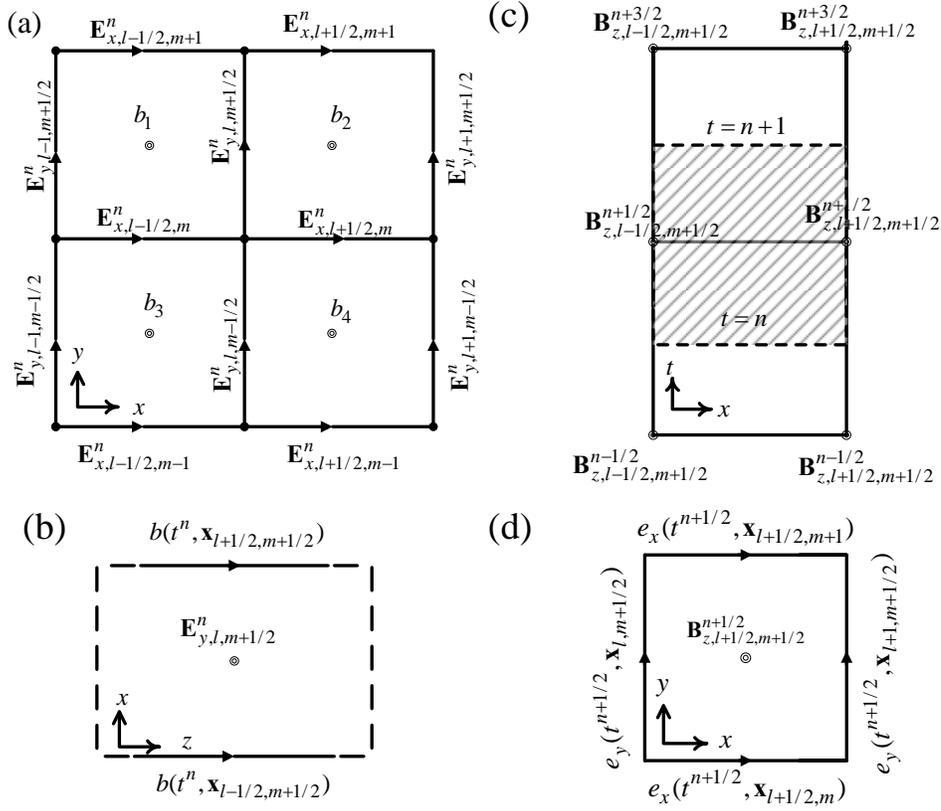

**Figure 4: (a)** x-y layout of four adjacent domain elements at $t = t^n$ showing space locations of four nodal values of **b**. Here values of **b** are not located in half integer time levels, therefore we have shown them by $b_1 = b_z(t^n, \mathbf{x}_{l-1/2,m+1/2})$, $b_2 = b_z(t^n, \mathbf{x}_{l+1/2,m+1/2})$, $b_3 = b_z(t^n, \mathbf{x}_{l-1/2,m-1/2})$, $b_4 = b_z(t^n, \mathbf{x}_{l+1/2,m-1/2})$. **(b)** x-z layout of an element at the same time as (a), space integration of Ampere's law is carried over this element. **(c)** x-t layout of two adjacent elements at $y = y_{m+1/2}$. Domain of time integration of Ampere's law is determined. **(d)** Another x-y layout at time $t = t^{n+1/2}$, space integration of Faraday's law is carried over this element.

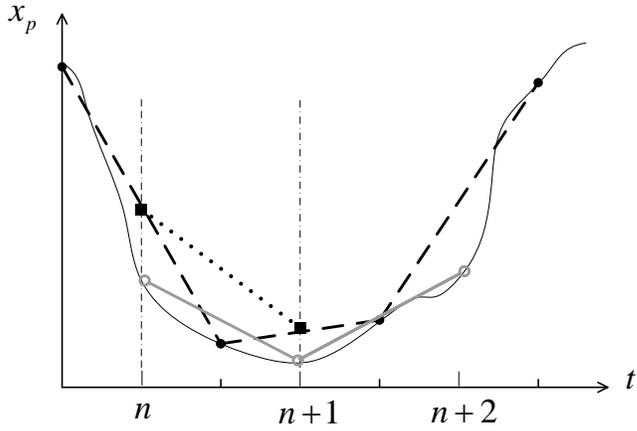

Figure 5: Black solid curve is a schematic of particle position as a function of time. The gray solid lines present commonly used linear interpolation, broken lines present our new interpolation and dotted line presents time average approximation to our interpolating curve.

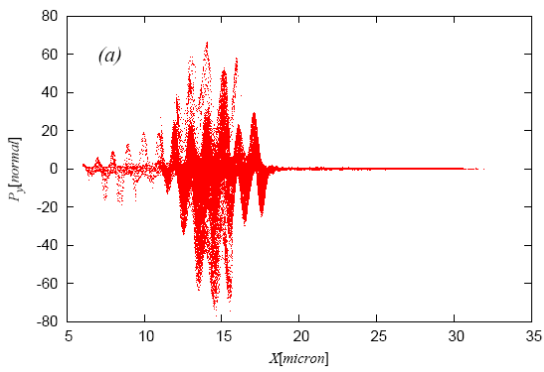
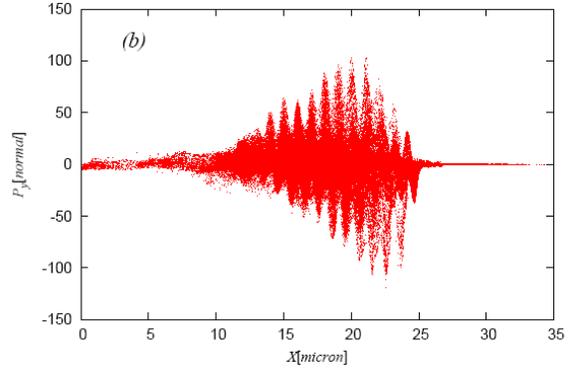
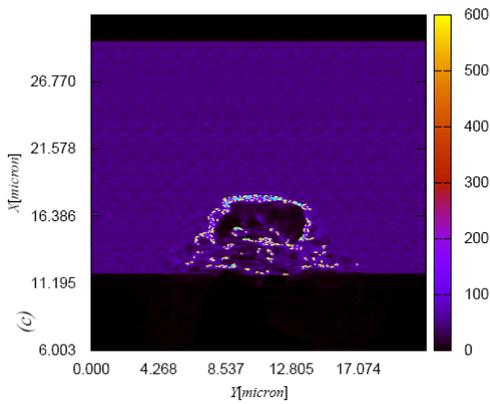
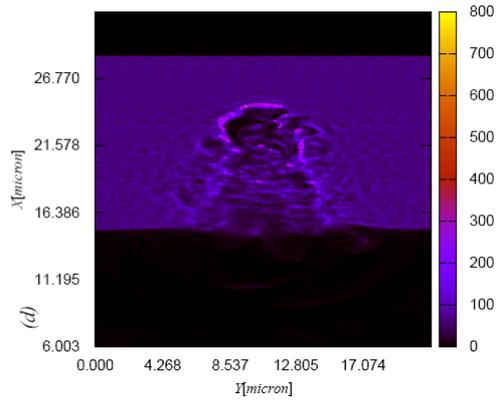

**Figure 6:** Visualization of laser-plasma interaction for parameters given in text. (a) $x - p_y$ (transverse) phase slice of electrons at $t = 30\,fs$ and (b) at $t = 60\,fs$, (c), (d) corresponding density maps of electrons at these times.

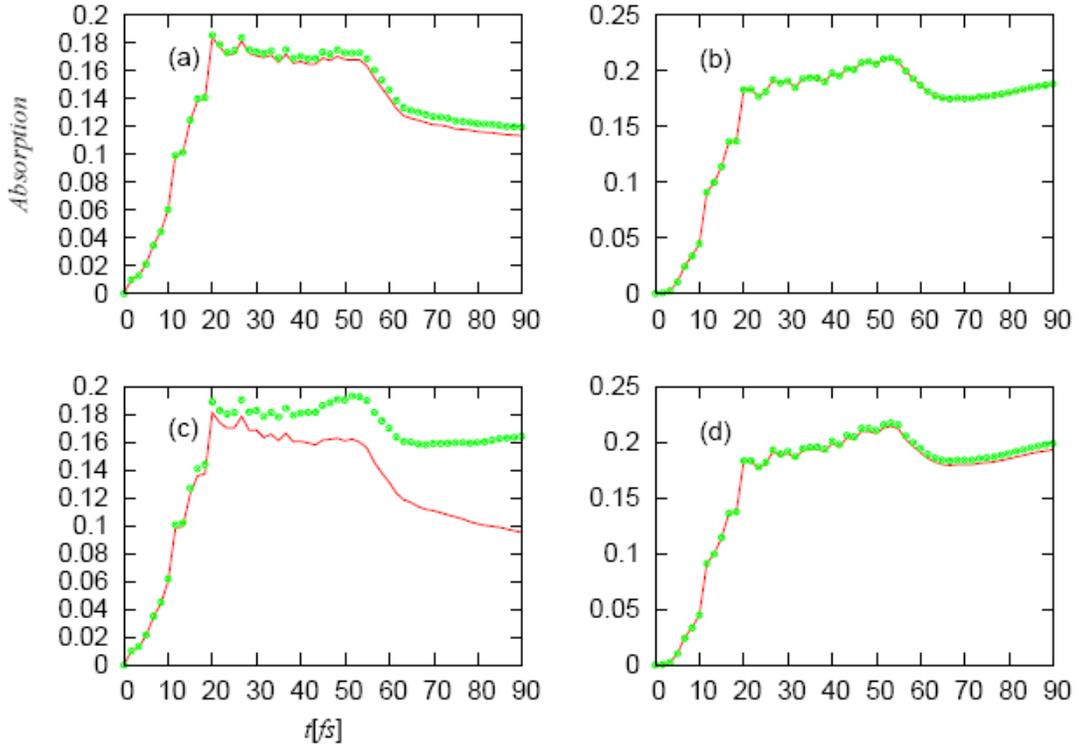

**Figure 7:** Absorption history for parameters in text. Line shows absorbed radiation and open circles show energy gain by particles. (a), (b) results of our simultaneous integration scheme for $N_{pc} = 0.5$ and $N_{pc} = 5$, (c) and (d) results of leapfrog scheme for the same $N_{pc}$s as (a) and (b) respectively.